\documentstyle[preprint,aps]{revtex}
\begin{document}

\newcommand\beq{\begin{equation}}
\newcommand\eeq{\end{equation}}
\newcommand\bea{\begin{eqnarray}}
\newcommand\eea{\end{eqnarray}}
\newcommand\bc{\begin{center}}
\newcommand\ec{\end{center}}
\draft
\title{Rotating fermions in two dimensions:   Thomas Fermi Approach}
\author{Sankalpa Ghosh$^1$, M.V.N.Murthy$^1$, Subhasis Sinha$^2$}
\address{1. The Institute of Mathematical Sciences, Chennnai 600113, India}
\address{2. Laboratoire Kastler-Brossel, Ecole Normale Superieure,  
75231 Paris Cedex 05. France} 
\maketitle

\begin{abstract}
Properties of confined mesoscopic systems have been extensively studied
numerically over recent years. We discuss an analytical approach to the
study of finite rotating fermionic systems in two dimension.  We first
construct the energy functional for a finite fermionic system within the
Thomas-Fermi approximation in two dimensions. We show that for specific
interactions the problem may be exactly solved.  We derive analytical
expressions for the density, the critical size as well as the ground state
energy of such systems in a given angular momentum sector. 
\end{abstract}
\vskip .5 true cm

\maketitle
\narrowtext
\newpage

\section{Introduction}

The observation of Bose-Einstein Condensation(BEC) in extremely cold
dilute gas of atoms has generated a lot of activity both on experimental
as well as theoretical front on the properties of trapped dilute gases.
There is now a renewed focus on the properties of trapped dilute gas of
fermionic atoms at low temperatures. A recent effort in this direction is
the experimental observation of quantum degeneracy in a dilute gas of
trapped fermionic atoms\cite{Demarco}.  Several theoretical papers have
studies the properties of trapped dilute gas of fermionic atoms. Butts and
Rokshar\cite{rokshar} have studied the momentum and spatial distribution
of the noninteracting system in the Thomas-Fermi approximation. The ground
state properties and the addition energy spectra of a two-dimensional
interacting fermi system has been studied by Sinha et al\cite{subhasis}.
Recently collective excitations of the system in the normal
phase\cite{brunn} and in the superfluid phase\cite{baranov} have also been
investigated. The thermodynamic properties of fermi gases including 
rotations has been analysed by Salasnich et al\cite{salasnich} 

Especially relevant to this paper is the recent interest in the properties
of rotating bose condensates\cite{Butts} and the experimental observation
of vortices in stirred BEC of $^{87} Rb$\cite{expt}. The properties of
ground and low excited states of a rotating, weakly interacting
Bose-Einstein condensate in a harmonic trap has been investigated recently
in various limits\cite{Mot1,kavoulakis}.  Ho and Ciobanu\cite{Ho} have
recently examined theoretically the nature the ground state of a
rotating trapped fermi gas in two and three dimensions in the 
noninteracting limit. They show that the density profile acquires 
features reflecting the underlying Landau level like energy spectrum. 
Properties of the spectrum of such a system had been earlier investigated 
by Bhaduri et al\cite{bhaduri}. 

In this paper we investigate a system of rotating fermions in two
dimensions confined within a parabolic potential with some model
interactions. We show that within the Thomas-Fermi approximation the
system may be solved exactly and discuss the consequent results. We follow
closely the formalism developed by Gallego and Das Gupta\cite{gdg} who
developed the Thomas-Fermi method for rotating nuclei. The attractive
feature of Thomas-Fermi method is the ease with which non-trivial
many-body solutions can be obtained even when they are interacting. As we
show later, in some cases the results may be obtained purely by analytical
methods. While this is not a full list of the exactly solvable models,
they are chosen for their relevance to physical systems.  The paper is
organised as follows. In section II we outline the derivation of the
Thomas Fermi energy for a system of interacting rotating fermions in
parabolic confinement. In section III we discuss the results for three
specific type of interactions between the fermions. We obtain analytic
expression and present numerical results for the interaction energy, the
rotational energy and the spatial density in a given angular momentum
sector for a given number of particles. The last section contains a
discussion of the results.

\section{Thomas Fermi Energy functional}

We derive the Thomas-Fermi(TF) energy functional for a confined two
dimensional rotating fermi gas starting from the microscopic Hamiltonian.
We shall follow the approach outlined by Gallego and Das gupta
\cite{gdg} who derived the TF density functional for rotating nuclei,
but applied to a two dimensional system.

The microscopic Hamiltonian of a two-dimensional rotating fermi gas 
confined in a harmonic potential is given by,
\bea
H &=& \sum_{i=1}^N \frac{{\bf p}_i^2}{2m} + \sum_{i<j}V(\vec 
r_i,\vec r_j) +\sum \frac{1}{2}m \omega_0^2 r_i^2 \nonumber \\
& &\mbox{}-\sum m\omega[{\bf r_i} \times {\bf p_i}], 
\label{MBH} 
\eea
where $V(\vec r_i, \vec r_j)$ denotes the two body interaction between 
fermions, $\omega_0$ and $\omega$ denote the confinement and rotational 
frequencies respectively.

We want to minimise the energy of this system subject to the constraints
\bea 
\int d \vec r \rho(\vec r) &=& N,  \label{cons1} \\
\int d \vec r d\vec p f(\vec r,\vec p)(\vec r \times \vec p) &=& L, 
\label{cons2} 
\eea  
where $N$ and $L$ are total particle number and the total Angular
momentum, $f(\vec r,\vec p)$ is the semiclassical phase space density
which we shall define later.  The configuration space density $\rho$ is
obtained by integrating the phase space density over the momentum space. 
 
Using Lagrange multiplier $\mu$ and $\omega$ for each constraint
respectively and regrouping the momentum dependent term we obtain the
following expression for the energy functional 
\bea 
E[\rho] &=& \int d\vec r d\vec pf (\vec r, \vec p)[\frac{p^2}{2m}
- \omega(xp_y - yp_x)] \nonumber \\ 
& & \mbox{} +  \int d{\vec r}[V(\vec r) - \mu \rho(\vec r)]+ \mu N
+ \omega L \label{enfun} 
\eea
where the mean-field one body interaction $V(r)$ is 
\beq 
V(r) = \frac{1}{2}m \omega_0^2 r^2 + V_{H}(\vec r).
\eeq
The Hartree term is given by,
\beq 
V_{H}(\vec r)=  \int d^2r V(\vec r, \vec r~') \rho(\vec r~'). 
\eeq
We ignore the exchange correction as it is in general subdominant 
compared to the Hartree term.

The semiclassical phase space density is given by 
\beq 
f(\vec r, \vec p) = \frac{2}{ (2\pi \hbar)^2}\Theta(\mu - \epsilon(\vec p,
\vec r)), 
\eeq 
where we have accounted for a factor of 2 due to spin degeneracy and 
\beq 
\epsilon(\vec p, \vec r) =\frac{p^2}{2m} +V(\vec r)
- \omega(xp_y - yp_x)
\eeq
denotes the energy density in phase space. We may rewrite this by
completing the square as,
\beq 
\epsilon(\vec p, \vec r) =\frac{1}{2m}[(p_x+m\omega y)^2 + (p_y -m\omega
x)^2] +V(\vec r)- \frac{1}{2}m\omega^2 r^2.
\eeq
The shifted momentum simply indicates that the center of the Fermi sphere
at any given point $\vec r$ in the rotating frame is displaced from the
usual $\vec p =0$. This shift is of no relevance within the classical TF
approximation except for the appearance of the extra centrifugal term in
the energy functional.

The TF energy functional is therefore given by,
\beq 
E[\rho] = \int d^2 r~[\frac{\pi \hbar ^2 \rho^2}{2m} + \frac{1}{2}m
\Omega r^2 + V_{H} - \mu] \rho(\vec r) + \mu N + \omega L 
\label{E}
\eeq
where
\beq 
\Omega^2~=~\omega_0^2 - \omega^2~=~ C\omega^2 
\eeq
denotes the effective frequency in the presence of rotations. 
The TF equation for the spatial density is obtained by a variation of the
energy functional, namely,
\beq 
\frac{\pi \hbar^2 \rho}{m} + \frac{1}{2}m\Omega^2 r^2 
+ V_{H}  =\mu 
\label{TFE} 
\eeq
To obtain the ground state density in TF approximation one has to solve
the above equation self-consistently with the boundary condition 
that the density vanishes beyound the classical turning point $r_0$, that 
is 
\beq \rho(r) = 0 , ~~~r \ge r_0. \eeq
Note that in the absence of rotations $\Omega=\omega_0$ and the solution
of the above equation describes the ground state in the absence of
rotations which has been analysed in detail before\cite{subhasis}.
We may remark that in general the above equation is difficult to solve
analytically except in specific cases which we shall discuss later.

The angular momentum carried by the system is given by,
\beq 
L = \int d^2 r d^2 p f (\vec r, \vec p)(xp_y - yp_x) 
=m\omega\int d^{2}r \rho(\vec r) r^2 
\eeq 
which is simply the classical expression for the angular momentum.
We may therefore identify the Lagranges multiplier $\omega$ with the
angular frequency.

We may also define the related energy functionals. The free energy of the
system is given by,
\beq 
F = E - \mu N - \omega L 
\eeq
The TF equation (\ref{TFE}) may also be obtained just by varying the free 
energy. We also define the energy in the rotating frame through 
\beq 
E = E' + \omega L. 
\eeq
where E is the total energy in the static frame.  With the help of eq.
(\ref{TFE}) it can be shown that irrespective of the nature of the
interaction $E'$ is given by,
\beq 
E' = \frac{1}{2}\mu N + \frac{1}{4} m \Omega^2 \int d^2r~ r^2 \rho.
\eeq
This concludes the basic TF formalism applied to rotating fermionic
systems. In the following section we solve the TF equation for specific
cases. Note that it is necessary that we choose $\omega \le \omega_0$ as
otherwise the system becomes unstable.

\section{ Some exactly solvable models}

We now discuss  special cases of interacting fermi gases which may 
be solved exactly. These models are chosen not only to facilitate analytical 
calculations, but also because in some limits these approximate the 
effective electron-electron interaction in a two-dimensional fermionic 
system. We note that the effective interaction may be dependent on the 
device characteristics as well as the number of particles. As such we 
find that it is useful to work with different approximate forms of 
interactions. Throughout we assume the system is spin unpolarized. The 
partially and fully polarized system in the presence of magnetic field 
will be discussed in a subsequent paper.

\subsection{Contact Interaction}
\label{contact}
The first and the simplest is the contact interaction. In general any
short-range interaction may be written in terms of a gradient
expansion\cite{Trug}.  The leading term in the expansion is always the
contact interaction of the form 
$$ V(\vec r) = V_0 \delta(\vec r),$$ 
where $\vec r$ denotes the relative coordinate between any two fermions.  
In a gradient expansion the factor $V_0$ is related to the leading moment 
of the potential\cite{Trug,murthy}. 

The free-energy is then given by,
\beq
F[\rho] = \int d^2r \big{[} \frac{\hbar^2 \pi \rho^2}{2m} + 
\frac{1}{2}
m\Omega r^2 \rho (\vec r) + \frac{V_0}{2} \rho^2] - \mu \int d^2 r 
\rho(\vec r) 
\eeq

The TF equation for the spatial density then becomes
\beq 
\frac{\hbar^2 \pi}{m}(1 + g)\rho(\vec r)+ \frac{1}{2}m \Omega^2
r^2 = \mu, 
\label{TF1} 
\eeq 
where $g = \frac{V_0 m }{\pi\hbar^2}$ which is dimensionless. Note that 
eq.(\ref{TF1}) is identical in form to the case when there is no 
rotation. Only the effective frequency $\Omega$ is different now.
The solutions are therefore obtained easily by merely replacing the 
confinement frequency by the effective frequency.  Since the density 
vanishes beyond a turning point $r_0$, the chemical potential is related 
to the turning point by,
\beq 
\mu = \frac{1}{2}m\Omega r_0^2. 
\eeq
The turning point and hence the chemical potential are therefore 
determined by the total number of particles $N$ which is fixed.

Substituting it back in the TF equation gives the expression for density
\beq 
\rho(r) = \frac{m^2 \Omega^2(r_0^2 - r^2)}{2 \pi \hbar^2(1 +{g})} 
\label{d1} 
\eeq.

The total number of particles is then obtained by integrating the
expression for density in eq.(\ref{d1})
\beq 
N ~=~ \int_0^{r_0} d^2r\rho(r) ~=~ \frac{m^2\Omega^2 r_0^4}
{4\hbar^2(1+{g})} 
\label{N1} 
\eeq
and the total angular momentum is given by
\beq 
L~=~ 
\int_0^{r_0} d^2r m\omega r^2 \rho(r) ~=~ \frac{m^3\omega
\Omega^2 r_0^6}{12 \hbar^2(1+{g})} \label{L1} 
\eeq
In a system where both $N$ and $L$ are fixed, the turning point $r_0$ and 
$\omega$ are determined by entirely by these two constraints. 

Using the expression for density and the total angular momentum $L$ 
in eq.(\ref{L1}) we can easily evaluate the expression of the energy.
Some straightforward algebra immediately yields,
\beq 
E = \hbar \omega_0 \sqrt{ L^2 + \frac{4}{9}(1+g) N^3} 
\eeq
Note that when $L=0$ and $g=0$ this is the standard result for TF energy of 
the ground state of a fermionic system in two dimension in an oscillator 
confinement. The effect of interaction is to scale the density and hence 
the energy by a factor which involves $(1+g)$ and the effect of rotations 
is contained in the $L$ dependent term. 

The interaction energy alone may be separated, by computing the 
interaction term in the density functional using the TF solution for the 
spatial density. We have therefore,
\beq  
E_{int} = \hbar\omega_0\frac{2N^3
{g}}{9\sqrt{L^2 +\frac{4}{9}N^3(1+{g})}}. 
\eeq
Note that with the increase in angular momentum the effect of interaction 
decreases as it should.

A few general remarks are in order here. In the actual many body system, 
there will be in general a tower of states in a given angular momentum 
sector. However, as is obvious from the expression for the total energy 
the TF method gives the lowest energy state or the ground state in each 
angular momentum sector. This is easily seen by taking the limit $L >> 
N$, when the energy is simply $\hbar \omega_0 L$ which is the energy 
obtained without any radial excitations. These states are often referred 
to as Yrast states in the literature.

We briefly discuss the case with an arbitrary but very short range 
interaction. The leading term, as remarked before, is simply the contact 
term discussed above. The corrections may be systematically worked out 
using the gradient expansion method\cite{Trug,murthy}. Below, we mention 
the results from the next to leading order correction. Again the 
corrections may be obtained analytically. The results are given as 
corrections to the results obtained with contact interaction alone.

The next higher order  correction to the contact interaction in the 
gradient expansion of a short range potential is of the 
form\cite{murthy},  
\beq 
V_{SR} =- V_2 \nabla^2 \delta (\vec r) 
\eeq
where $V_2$ is related to the second moment of the interaction potential. 
Note that in the gradient expansion only even moments appear in each term. 
The Thomas Fermi free energy functional is given by 
\beq
F[\rho] = \int d^2r \big{[} \frac{\hbar^2 \pi \rho^2}{2m} +
\frac{1}{2}
m\Omega r^2 \rho (\vec r) + \frac{V_0}{2} \rho^2 - \frac{V_2}{2}
\nabla^2 \rho] - \mu
\int d^2 r
\rho(\vec r)
\eeq

Variation with respect to density gives the following self consistent 
equation for the density
\beq -V_2 \nabla^2 \rho(\vec r) + \frac{\hbar^2 \pi}{m}(1 + g)\rho(\vec
r)+ \frac{1}{2}m \Omega^2 r^2 = \mu,
\label{TF2} \eeq
It can be easilly seen that for $V_2 = 0$ eq. (\ref{TF2}) reduces to 
eq. (\ref{TF1}).

The circularly symmetric solutions of the above inhomogeneous partial 
differential equation can be found easily. Replacing $V_2$ with the 
dimensionless coupling $g'$ , where $ g' = \frac{V_2 m^2 \omega_0}{\pi 
\hbar^3}$ we can write the solution as 
\beq \rho  = \rho_c + \rho_{cr} \eeq
where $\rho_c$ is given by eq.(\ref{d1}) with the contact term alone  
and  $\rho_{cr}$ is the extra term which can be written  in terms of
modified Bessel functions
\beq  \rho_{cr} =
\frac{2}{g'k^4}
(1-\frac{\omega^2}{\omega_0^2})
[\frac{I_0(kx)}{I_0(kx_0)} -1] \eeq
where 
\beq k^2 = \frac{1+g}{g'} \eeq
Here we have  used
the dimensionless variable $x=\frac{r}{l_0}$ with oscillator length scale 
$l_0 = \frac{\hbar}{m \omega_0}$ as the unit of length. Here after we use 
the subscript "$c$" to denote the contribution from the contact term alone. 

The corrections to various other relevant quantities may be easily found 
with the density. The total  no. of particles N can be similarly given by 
the following expression. 
\beq N ~=~ N_c +\frac{2}
{g'k^4}~(1-\frac{\omega^2}{\omega_0^2})~[ 
\frac{2I_1(kx_0)}{kI_0(kx_0)} - x_0^2] 
\eeq 
The total angular momentum $L$ ican also be written as 
\beq 
L = L_c + \frac{4\hbar}{g' k^4}
(1-\frac{\omega^2}{\omega_0^2})
\frac{\omega}{\omega_0}~
[\frac{1}{k}~[x_0(x_0^2+4)\frac{I_1(kx_0)}{I_0(kx_0)} -2x_0^2]
-\frac{x_0^4}{4}] \eeq
The  energy is again given by
\beq 
E = \frac{\mu N}{2}+[1+\frac{\omega_0^2}{4\omega^2}
(1-\frac{\omega^2}{\omega_0^2})]\omega L 
\eeq 

The interaction energy is given by
\beq E_{int} = E -
\frac{1}{2}(2+\sqrt{(1-\frac{\omega^2}{\omega_0^2})})
\omega L - \hbar\omega_0 \int d^2r\rho^2. 
\eeq
We note that the last term which involves the second moment of the 
density can be expressed in terms the modified Bessel functions.

\subsection{Logarithmic Interaction}

We now consider the case where two fermions interact via a logarithmic
interaction. For quantum dot systems this is close to being realistic
interaction at short distances where as the interaction is varies
inversely as the distances at long distances\cite{DZ}. The two dimensional
TF atom with logarithmic interactions was solved earlier by Bhaduri {\it
et. al.}\cite{bhaduri2} and was applied to quantum dots recently by Sinha
{\it et. al.}.\cite{subhasis}) in order to explain the shell effect in
quantum dots. The extension of these results for the rotating case is
quite staright forward. We give the main results below:

The TF mean field is given by
\beq 
V(r) = \frac{1}{2}m\Omega^2 r^2 - e_f^2\int d^2 r'
\rho(\vec r')ln(\frac{|\vec r - \vec r'|}{a}),
\eeq
where $a$ is an arbitrary parameter with the dimension of length. In order
that the interaction is repulsive it is essential that the parameter $a$
is much larger than any other length scale in the problem.  The largest
length scale in the problem is the turning point and we set $a=r_0$. While
this may seem arbitrary, in the analysis of the ground state of many
fermions this choice is justified by a numerical analysis of the data on
addition spectrum\cite{subhasis}.  The effective strength of the
interaction is denoted by $e_f^2$ . Unlike in the three-dimensional
Coulomb case, this parameter is not dimensionless, but has the dimensions
of energy. The only change from the analysis of the ground state in the
absence of rotations is in the first term where the effective frequency is
different from the confinement frequency. We therefore only give essential
results below and details may be looked up in Ref.\cite{subhasis}. 

The spatial density is given by the solution of the differential equation
\beq 
\frac{\hbar^2 \pi}{2m}\nabla^2\rho(\vec r) = \pi e_f^2 \rho(\vec r)
-2 m \Omega^2 
\eeq 

For a circularly symmetric spatial distribution, the above equation is
easily solved\cite{subhasis}. The solutions are given by the modified
Bessel function as,
\beq
\rho(x)=\frac{2b^2}{l_0^4}\frac{1}{(1-\frac{\omega^2}{\omega_0^2})}
[1 - \frac{I_0(x)}{I_0(x_0)}], 
\label{d2}
\eeq
where we have again introduced the dimension-less variables,
\beq
x=\frac{r}{b} ~~~~ x_0 = \frac{r_0}{b}
\eeq
with $b = \frac{\hbar^2}{2m e_f^2}$ has the dimension of length

with $x_0$ is again the classical turning point in units of $b$.

The total particle number can then be obtained by integrating this density
\beq 
N ~=~ \int_0^{2\pi}\int_0^{r_0}d{\bf r}\rho(\vec r) ~=~
\frac{4b^4}{l_0^4}\frac{1}{(1-\frac{\omega^2}{\omega_0^2})}
[\frac{x_0^{2}}{2} - \frac{x_0 
I_1(x_0)}{I_0(x_0)}] 
\eeq 
where $l_0^2=\frac{\hbar}{m\omega_0}$

Since $N$ is fixed, the turning point  $x_0$ can be determined from 
this equation. Similarly one can calculate the total angular momentum. After
some staright forward algebra it can be written as
\beq 
\frac{L}{\hbar} = \frac{b^2}{l_0^2}\frac{\omega}{\omega_0}[N(x_0^2 +4) -
\frac{b^4x_0^4}{l_0^4} 
\frac{\omega^2}{\omega_0^2} 
(1-\frac{\omega^2}{\omega_0^2})] 
\eeq
The energy is given as
\beq 
E ~=~ \mu\frac{N}{2} + [1+\frac{1}{4}
\frac{\omega^2}{\omega_0^2} 
(1-\frac{\omega^2}{\omega_0^2})] \omega  L 
\eeq
Therefore at a given $N$ and $L$ the constants $r_0$ and $\omega$ are 
determined by the above set of equations. This is in general done 
numerically unlike in the case of contact interaction in order to express 
the energy as a function of $N$ and $L$.
The interaction energy also can be written in a closed form and is
given by the following expression.
\beq 
\frac{E_{int}}{\hbar \omega_0} = \frac{b^2}{l_0^2}
(1-\frac{\omega^2}{\omega_0^2})
[x_0^2 -2N 
-2\frac{b^4}{l_0^4}
(1-\frac{\omega^2}{\omega_0^2})
\frac{I_1^2(x_0)}{I_0^2(x)}]
\eeq

\section{Discussion of Results}

We have outlined an analytical procedure based on the simple Thomas-Fermi 
approach to the many body problem of rotating fermions in two dimensions.
We have derived the density, the angular momentum as well the energy for 
model two body interactions. For simplicity we have chosen the contact 
interaction and logarithmic interaction. While these may not be the exact 
form of the interaction in the two dimensional systems they may under 
different circumstances approximately reflect the true nature of the 
interactions. 

We further illustrate these analytical results with numerical computation
of quantities of interest by taking $N=50$ which is reasonably large for
TF results to be valid. In Fig. 1 we have plotted the change in angular
momentum as a function of rotational frequency measured in units of the
confinement frequency. The curves represent the behaviour of $L$ with the
two model interactions. While in the case of contact interaction the
relationship is almost linear all the way upto 0.6, the non-linear
dependence of $L$ on $\omega$ builds up much earlier in the case of
logarithmic interaction. From eq.(\ref{L1}) and eq.(\ref{N1})it is easy to
see that $$ \omega^2/(1-\frac{\omega^2}{\omega_0^2}) \propto L^2/N^3, $$ in
the case of contact interaction. Hence the nonlinearity sets in only when
the ratio between rotational frequency and the confinement frequency is
considerable.  

As seen in Figs. 2 and 3, the total and interaction energy are sensitive
to the behaviour of $L$ as a function of $\omega$. We have shown the ratio
of total energy with the same in the absence of rotations in Fig.2. As
expected the energy increases slowly at first, but becomes linear at high
angular momentum since $\omega L$ term starts dominating.  This is a
precursor of the expected behaviour in the Yrast region of the rotational
spectrum which we have discussed earlier. It will be instructive at this
point to mention the behaviour of the rotating nuclei in the Yrast region.
There at very high angular momentum sector though the excitation energy is
quite high, the nucleus remains cold as most of the energy is spent in
rotating the nucleus. In the Yrast region the interaction is of lesser 
importance as can be seen from Fig. 3. 

A related question of interest that is relevant to a rotating nuclei as 
well as the two dimensional fermionic system is that
under heavy internal stress generated by the centrifugal and Coriolis 
forces generated in the Yrast region (\cite{Bohr}) the density profile
may drastically change. This issue has been investigated by Gallego
and Das Gupta (\cite{gdg}) and requires the consideration of a 
non radially symmetric density. This question as well as system quantum 
corrections will be addressed in a forthcoming paper\cite{Sinha2}.

\begin{figure}
\label{fig1}
\caption{  Plot of the angular momentum versus the ratio
 $\frac{\omega}
{\omega_0}$.The dotted line is the result for contact interaction,
while the solid line gives the result 
for logarithmic interaction. The number of particles is kept same for all 
the cases at 50. The angular momentum is in units of $\hbar$}.
\end{figure}
 
\begin{figure} 
\label{fig2} 
\caption{  Plot of the total energy $E$ (scaled by $E$ at $L=0$)
 versus angular momentum $L$. Here also the solid line is for
logaritmic interaction while the dotted  line is for  contact
interaction. 
All other parameters are kept same as in Fig.1}
\end{figure} 

\begin{figure}
\label{fig3}
\caption{ Plot of the 
Interaction energy ($E_{int}$) again scaled 
by the $E_{int}$ at $L=0$ as a function of 
the angular momentum ($L$).
Here also the solid line is for logaritmic interaction while the dotted
line is for contact interaction.
All other parameters are kept same as in Fig.1}
\end{figure}


\begin{references} 
\bibitem{Demarco} B.Demarco and D. S. Jin, Science {bf 285}, 1703 (1999) 
 \bibitem{rokshar} D. A. Butts and D. S. Rokshar, Phys. Rev. {\bf A
55}, 4346 (1997)
 
\bibitem{subhasis}  Subhasis Sinha, R. Shankar, and M.V.N. Murthy, 
Phys. Rev. {\bf B 62},10896 (2000)

\bibitem{brunn} G. M. Brunn and C. W. Clark, Phys. Rev. Lett. {\bf 83} 
5415 (1999)

\bibitem{baranov} M. A. Baranov and D. S. Petrov, Phys. Rev. {\bf A 62}, 
041601(R) (2000)

\bibitem{salasnich} L. Salasnich, B. Pozz, A. Parola and L. Reatto, J. 
Phys. B: At. Mol. Opt. Phys. {\bf 33}, 3943 (2000)

\bibitem{Butts} D. A. Butts and D. S. Rokhsar, Nature {bf 397},327 (1999)

\bibitem{expt} K. W. Madison, F. Chevy, W. Wohlleben and J. Dallibard,
Phys. Rev. Lett. {\bf 84},806 (2000)

\bibitem{Mot1} B. Mottelson Phys. Rev. Lett. {\bf 83},2695 (1999)

\bibitem{kavoulakis} G. M. Kavoulakis, B. Mottelson and C. J. Pethick, 
cond-mat/0004307; A. D. Jackson, G. M. Kavoulakis, B. Mottelson and S. M. 
Reimann, cond-mat/0004309

\bibitem{Ho} T.L.Ho and C. V. Ciobau, cond-mat/0005508

\bibitem{bhaduri} R. K. Bhaduri, S. Li, K. Tanaka and J. C. Waddington, 
J. Phys. {\bf A 27}, L553 (1994)

\bibitem{gdg} J. Gallego and S. Das Gupta, Phys. Lett. B {\bf 270}, 
6 (1991).

\bibitem{Bohr} A. Bohr and B. R. Mottelson, {\it Nuclear
Structure},Vol. 2, (N. A. benjamin Inc., 1975)

\bibitem{Trug} S. A. Trugman and S. Kivelson, Phys. Rev. B, {\bf 31},5280
(1985)

\bibitem{murthy} R. K. Bhaduri, M. V. N. Murthy and M. K. Srivastava, 
Phys. Rev. Lett. {\bf 76}, 165(1996)

\bibitem{bhaduri2} R. K. Bhaduri, S. Das Gupta and S. J. Lee,  Am. J. Phys. 
{\bf 58},983 (1990)

\bibitem{DZ} F.C.Zhang and S. Das Sarma, Phys.Rev.B,{\bf 33}, 2903 (1986)  

\bibitem{Sinha2} Subhasis Sinha, Sankalpa Ghosh and  M. V. N. Murthy
(in preparation). 

\end{references}
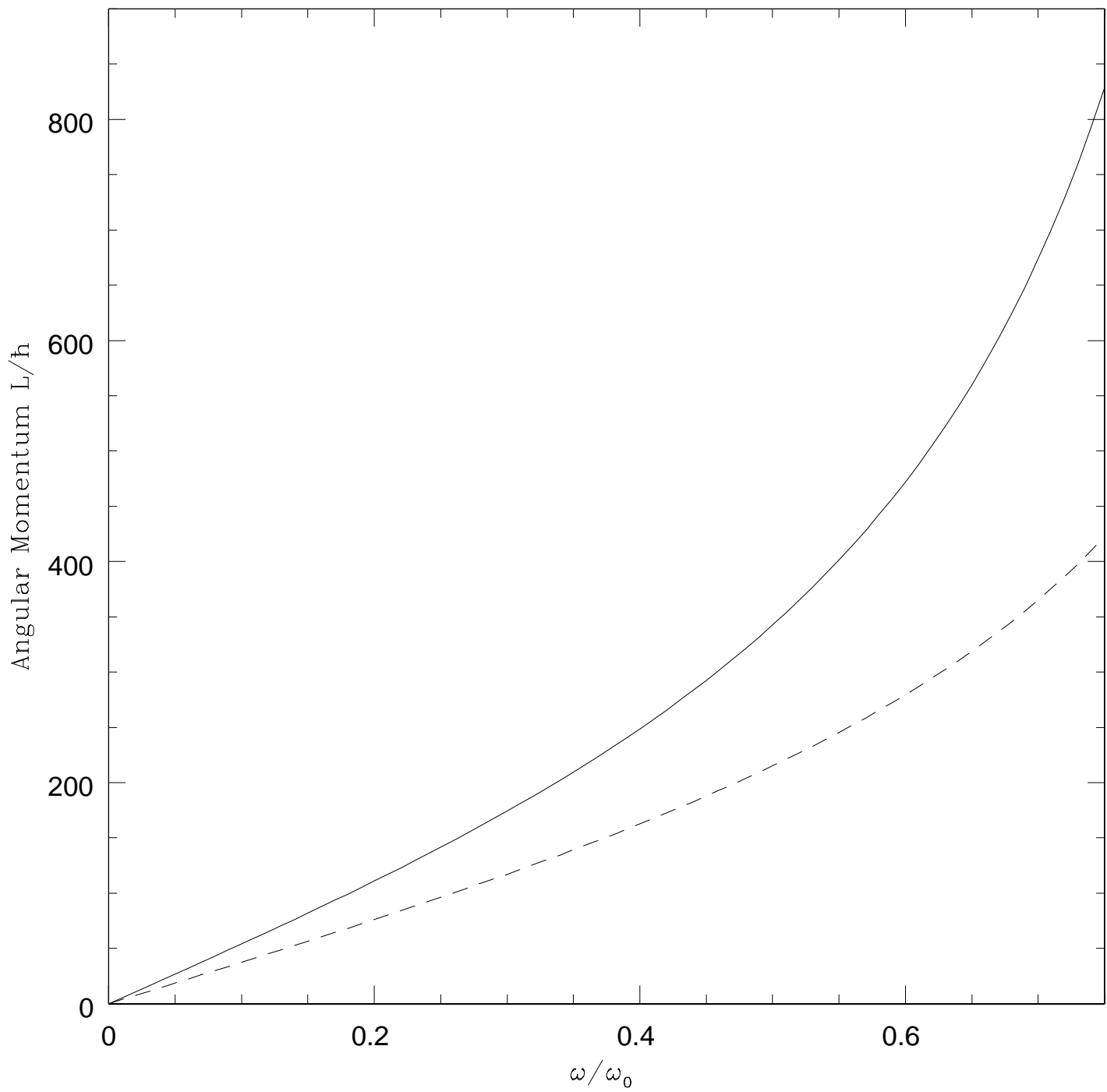
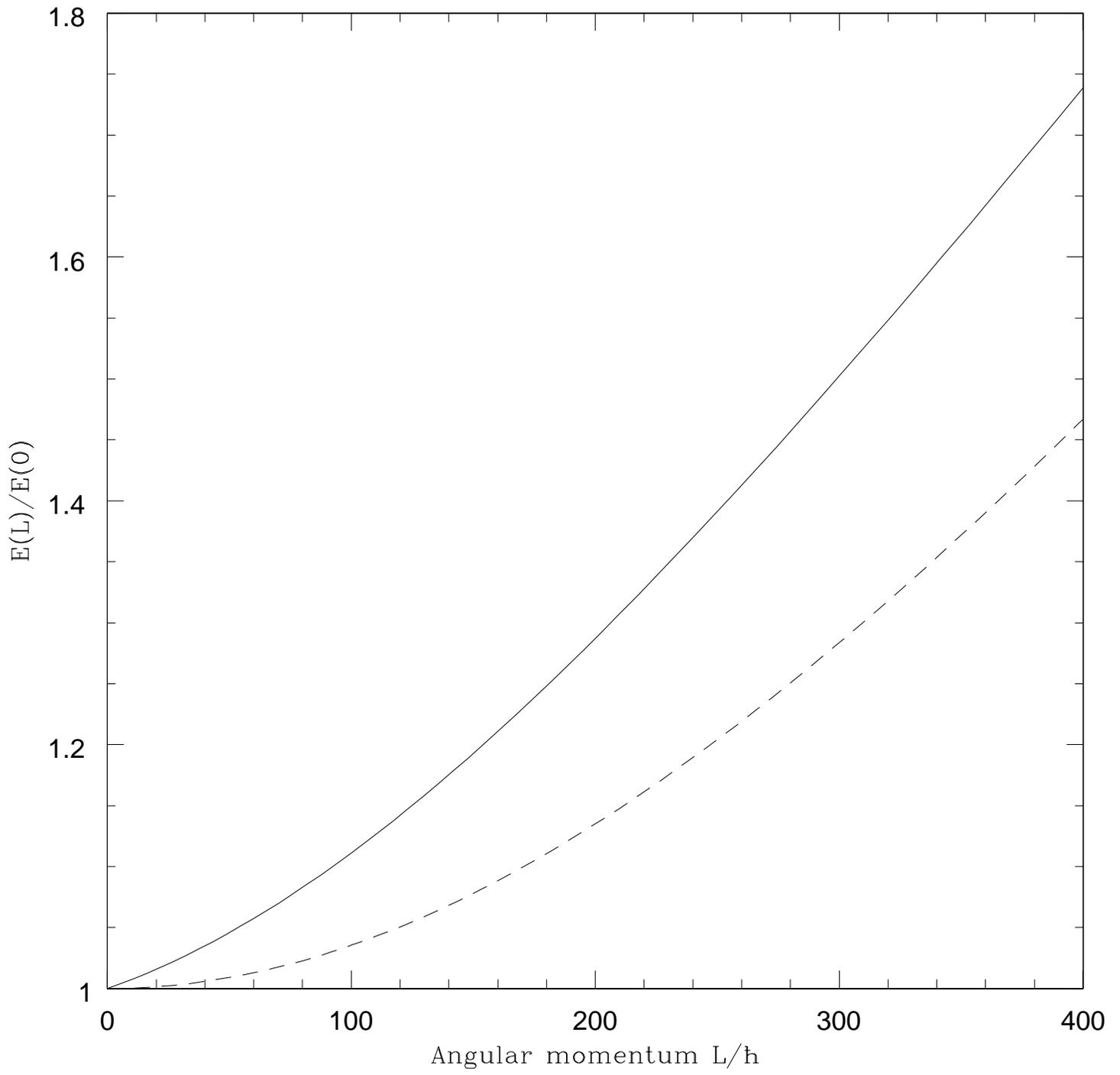
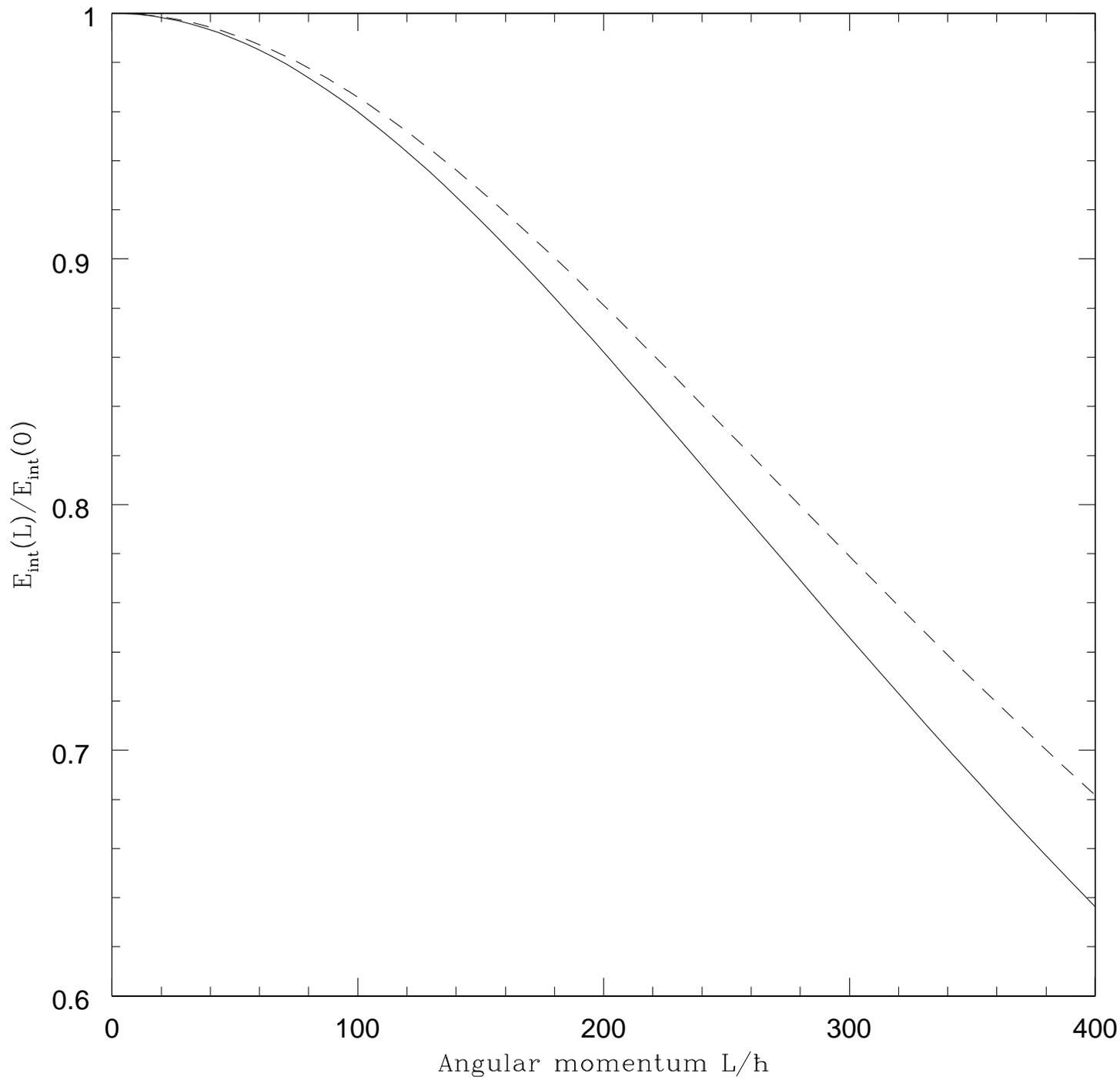
\end{document}